\documentclass[%
reprint%
,twocolumn
,aip%
,a4paper,superscriptaddress,nofootinbib,showkeys]{revtex4-1}
\usepackage{stdpreamble}

\usepackage{tikz}
\usetikzlibrary{matrix}

\newcommand{\lie}{\ensuremath{\mathsterling}}
\newcommand{\Diff}{\text{Diff}}

\begin{document}
\title{Rigidity of MHD equilibria to smooth incompressible ideal
  motion near resonant surfaces}
\author{David Pfefferlé}
\author{Lyle Noakes}
\affiliation{The University of Western Australia, 35 Stirling Highway, Crawley WA 6009, Australia}
\author{Yao Zhou}
\affiliation{Princeton Plasma Physics Laboratory, Princeton, NJ 08543, USA}
  
\date{\today}

\begin{abstract}
In ideal MHD, the magnetic flux is advected by the plasma motion, freezing flux-surfaces into the flow. An MHD equilibrium is reached when the flow relaxes and force balance is achieved. We ask what classes of MHD equilibria can be accessed from a given initial state via smooth incompressible ideal motion. It is found that certain boundary displacements are formally not supported. This follows from yet another investigation of the Hahm--Kulsrud--Taylor (HKT) problem, which highlights the resonant behaviour near a rational layer formed by a set of degenerate critical points in the flux-function. When trying to retain the mirror symmetry of the flux-function with respect to the resonant layer, the vector field that generates the volume-preserving diffeomorphism vanishes at the identity to all order in the time-like path parameter.
\end{abstract}


\maketitle

\section{Introduction}
\label{sec:intro}
Renewed interest in stellarator design has sparked questions on the existence and accessibility of three-dimensional magneto-hydrodynamic (MHD) equilibria with ``good'' nested flux-surfaces~\cite{rosenbluth-1973,boozer-2010,white-2013,loizu-2015,loizu-helander,zhou-2019}.
Several numerical tools exist to obtain three-dimensional MHD equilibria by means of variational principles~\cite{vmec,spec,zhou-2014}, initial value problems~\cite{hint,hint2}, iterative methods~\cite{pies,siesta}, metriplectic formulations~\cite{bressan-2018}, analytic expansions around a given magnetic axis~\cite{landreman-sengupta-2018,landreman-sengupta-plunk}.
These methods aspire to produce and optimise the magnetic fields so that the field-lines lie on toroidally nested flux-surfaces~\cite{stellopt,stellopt2}, which is the basis of plasma confinement in magnetic fusion devices such as tokamaks and stellarators~\cite{helander-2014}.
 
Restricted to flux-surfaces, the magnetic field is an integrable Hamiltonian vector field~\cite{morrison-greene,holm-marsden-ratiu-weinstein,cary-littlejohn}.
Under the assumption of translational and/or rotational symmetry (isometries), the MHD equilibrium problem reduces to a two-dimensional elliptic PDE for the scalar flux-function, called the Grad--Shafranov equation \cite{shafranov-1963,grad-1967}. Flux-surfaces naturally correspond to the level sets of the flux-function extruded to surfaces along the direction of symmetry.

The question is whether an initial configuration with nested flux-surfaces can be smoothly deformed through a family of MHD equilibria to reach a target three-dimensional configuration with equivalent (diffeomorphic) flux-surfaces. The answer is in general no; the deformation of the boundary must be chosen very carefully. In the language of magnetic confinement fusion, the issue is that flux-surfaces with periodic field-lines (rational rotational transform) are sensitive to resonant perturbations. 

The main reason smooth deformation of flux-surfaces is generally not possible is illustrated in this paper by proving nonexistence of smooth solutions to the so-called Hahm--Kulsrud--Taylor (HKT) problem \cite{hahm-kulsrud} for a given class of boundary perturbations. Known methods of solutions~\cite{hahm-kulsrud,rosenbluth-1973,boozer-2010,dewar-2013,zhou-2016,zweibel-li} produce either discontinuities, an infringement of force-balance, boundary layers, or finite resistivity.

The paper is organised as follows. In section \ref{sec:grad-shafranov}, the MHD equilibrium problem is set up in the simpler case of a slab with an ignorable coordinate. In section \ref{sec:diffgeo}, elements of differential geometry are introduced to show that the force-balance condition can be cast as the vanishing of a Poisson-bracket between the flux-function and its Laplacian. In section \ref{sec:idealmotion}, the frozen-in condition of ideal MHD is identified with the precomposition of the flux-function by a diffeomorphism (pull-back). The requirement for smooth incompressible ideal motion naturally follows from the preservation of the Poisson-bracket (symplectomorphism) as well as Faraday's law of induction. The idea is then to generate a family of diffeormorphisms via a sequence of $t$-dependent Eulerian velocity fields in order to smoothly map an initial equilibrium state into another (isotopy). Examples of one-parameter subgroups of diffeomorphisms are shown to illustrate the challenge of retaining force-balance beyond linear order in the $t$-parameter. In section \ref{sec:hkt}, we prove that there are in fact no smooth Eulerian velocity fields that can deform the level sets of the initial flux-function of the Hahm-Kulsrud-Taylor and conform with a mirror-symmetric boundary condition. With this result, we conclude in section \ref{sec:conclusion} that MHD equilibria with a line of degenerate critical points in the flux-function, i.e. rational surfaces, are rigid to large classes of smooth incompressible ideal deformations.

\section{The Grad--Shafranov equation in a slab}
\label{sec:grad-shafranov}
The resonant behaviour of magnetic field-lines near rational surfaces is a known phenomenon that leads to magnetic island formation in toroidally confined plasmas such as in the tokamak and stellarator devices. We reduce the complexity of the discussion due to geometry by working in the simpler setting of a slab, $(x,y,z)$, where $x$ plays the role of a \emph{radial variable}, $y$ the periodic \emph{poloidal angle} and $z$ the ignorable \emph{toroidal angle}. We demonstrate that a core obstruction to accessing equivalent equilibrium states through ideal motion already arises in the flat slab metric, which will carry over to the realistic toroidal geometry by elliptic regularity.

By symmetry, the form of the vector potential is, up to a gauge term, $\bm{A}=A_y(x,y)\nabla y + \Psi(x,y) \nabla z$ and the magnetic field $\bm{B}=I(x,y)\nabla z + \nabla \Psi \times \nabla z$. Throughout the paper, $\Psi(x,y)$ will be called the \emph{flux-function}.
An MHD equilibrium without flow is a configuration such that $\bm{J}\times\bm{B} = \nabla p$, where $p(x,y)$ is the plasma pressure and $\bm{J}=\curl\bm{B}$ is the current density. Using the above representation, MHD force-balance is achieved when the gradient of the pressure $p(x,y)$ and the gradient of the longitudinal current (or guide field) $I(x,y)$ are collinear with the gradient of the flux-function, $\nabla p\times \nabla \Psi=0$ and $\nabla F\times \nabla \Psi=0$, i.e. $\nabla p = p'\nabla \Psi$ and $\nabla I=I'\nabla \Psi$. Then, the problem reduces to the Grad--Shafranov equation
\begin{equation}
  \label{eq:gs}
\Delta \Psi = II'+p'=V'(\Psi),
\end{equation}
where $\Delta = \nabla\cdot \nabla$ is the Laplacian operator. This formulation of the MHD equilibrium problem is similar to that of reduced MHD~\cite{strauss-1976,krauss-2016}. The only difference is that the guide-field in reduced MHD is constant, $I(x,y)=B_0$, whereas in our setting, this component is an unspecified function of the flux, $I(x,y)=I(\Psi(x,y))$.

A convenient way to rephrase the force-balance condition in order to eliminate the arbitrary pressure and current functions is
\begin{equation}
  \label{eq:force-balance-cartesian}
  \curl(\Delta\Psi \nabla \Psi)=0.
\end{equation}
\begin{remark}
  The Grad--Shafranov equation (\ref{eq:gs}) implies equation (\ref{eq:force-balance-cartesian}). The converse is in general not true, e.g. $\Delta \Psi$ piecewise-constant.
\end{remark}
A flux-function $\Psi$ that satisfies equation (\ref{eq:force-balance-cartesian}) determines an MHD equilibrium. For example, harmonic functions are special cases describing vacuum fields, $\Delta\Psi=0$. Solutions to the uniform current density or constant pressure-gradient Poisson equation, $\Delta \Psi = \const$, form another particular class, comprising the initial configuration of the HKT problem, $\Psi_0(x,y)=\frac{1}{2}(1-x^2)$. Taylor-relaxed fields~\cite{taylor-1974} are a third kind of solutions satisfying the Helmholtz equation $\Delta \Psi = -\mu^2 \Psi$, and coincide with the eigenfunctions of the Laplacian.

\section{Differential geometric setting}
\label{sec:diffgeo}
Useful information and structure is obtained by rephrasing the problem in the language of differential geometry. The advantage is that part of the conclusions will be independent of the specific choice of coordinates and straightforwardly transposable to more complicated geometries such as the toroidal case, as well as three-dimensional equilibria.

Briefly, let $M=\R \times S^1 = \{(x,e^{iy}) | x\in\R, y\in\R\}$. We equip $M$ with the standard volume form $\omega=dx\wedge dy$ and the standard Euclidean metric $\langle X,Y\rangle(p) = X^x(p)Y^x(p)+X^y(p)Y^y(p)$, where $Xf(p)=X^x(p)\partial_x f(p) + X^y(p)\partial_y f(p)$ represents the vector field $X$ acting as a differential operator on the smooth function $f:M\to \R$. The interior product (contraction) of a vector field $X$ and the volume form $\omega$ is defined by $[i_X\omega](Y)=\omega(X,Y), \forall Y\in \Gamma(TM)$. The so-called \emph{musical} isomorphism between vector fields and one-forms is established through the Riemannian metric (relation between co-variant and contra-variant tensor fields), e.g. the \emph{flat} of a vector field is the unique one-form such that $\langle X , Y \rangle = X^\flat(Y), \forall Y\in \Gamma(TM)$. The Hodge star $\star$ operator, also based on the Riemannian metric, is an isomorphism between $k$-forms and $n-k$ forms on $M$ with the property that for $\alpha,\beta \in \Omega^k(M)$, $\alpha\wedge \star\beta = \langle \alpha, \beta\rangle \omega$.

\begin{remark}
  We will identify $dy$ as the closed but misleadingly not exact one-form on $S^1$ such that $\oint dy = 2 \pi m$ where the integer $m\in \mathbb{Z}$ depends only the closed integration path.
\end{remark}

The volume form $\omega$ plays the role of a symplectic form on this two-dimensional manifold, enabling the identification of \emph{Hamiltonian vector fields}: given a smooth function $F:M\to \R$, there is a unique vector field $X_F$ such that
\begin{align}
  i_{X_F}\omega &= dF, & X_F^\flat = -\star dF.
\end{align}
Hamiltonian vector fields are divergence-free, $(\div X_F)\omega:=\lie_{X_F}\omega = di_{X_F}\omega=d^2F= 0$.  This property means that $X_F$ is the generator of a one-parameter family of volume-preserving diffeomorphisms on $M$ (simultaneously symplectomorphisms), where the volume form is preserved (advected) along the flow of $X_F$. In addition, since $\lie_{X_F}F=dF(X_F) = i_{X_F}dF = i_{X_F}i_{X_F}\omega = 0$, the Hamiltonian vector field $X_F$ is tangential to curves of constant $F$.

As differential operators, Hamiltonian vector fields can be used to define an anti-symmetric bilinear operation on smooth functions $F$ and $G$ by
\begin{align}
  \{F,G\}
  := -\lie_{X_F}G
  = \omega(X_F,X_G)
  = \star (dF\wedge dG),
\end{align}
which is the same as $\{F,G\}\omega = dF\wedge dG$ in our
two-dimensional setting. This product rule qualifies as a \emph{Poisson
  bracket}; it satisfies the Leibniz rule,
\begin{align}
  \{FG,H\}
  = \{F,H\}G+ F\{G,H \},
\end{align}
as well as the Jacobi identity (see appendix~\ref{sec:proof-jacobi}),
\begin{align}
  \{F,\{G,H\}\} + \{G,\{H,F\}\} + \{H,\{F,G\}\} = 0.
\end{align}
In local coordinates, the Poisson bracket is simply computed as
\begin{align*}
 \{F,G\} = \partial_xF\partial_yG-\partial_yF\partial_xG = (\nabla F \times \nabla G) \cdot \nabla z.
\end{align*}

In this context, the flux-function $\Psi:M\to \R$ is the Hamiltonian for the \emph{poloidal (or helical) magnetic field} $B$, obtained via $i_{B}\omega = d\Psi$ or $B^\flat = -\star d\Psi$. This construction makes the magnetic field tangential to curves of constant $\Psi$, called \emph{flux-levels}. We will refer to the quantity $d\Psi$ as the \emph{magnetic one-form}.
\begin{remark}
  In local coordinates $B = \partial_y\Psi \partial_x-\partial_x\Psi\partial_y$ or equivalently $\bm{B}=\curl(\Psi\nabla z) = \nabla\Psi\times \nabla z$ is the poloidal (or helical) component of the magnetic field in the 3D slab picture.
\end{remark}

The force-balance condition of equation (\ref{eq:force-balance-cartesian}) is identified as the following property satisfied by the flux-function
\begin{equation}
  \label{eq:force-balance-diff}
  d(\Delta \Psi d\Psi) = d(\Delta \Psi) \wedge d\Psi = 0
  \iff \{\Delta \Psi, \Psi\} = 0
\end{equation}
where $\Delta = \delta d + d\delta = (\delta + d)^2$ is the Laplace-de Rham operator and $\delta=-\star d\star $ is the codifferential. Another way of viewing equation (\ref{eq:force-balance-diff}) is $\lie_B\Delta \Psi = 0$, namely that the magnetic field is tangential to the level sets of the Laplacian of the flux-function.




\section{Frozen-in condition and smooth incompressible ideal motion}
\label{sec:idealmotion}
An MHD equilibrium $\Psi$ is said to be accessible through ideal motion from an initial MHD equilibrium $\Psi_0$ if there exists a diffeomorphism (smooth map with smooth inverse) $\varphi:M\to M$ such that the following diagram commutes
\begin{center}
  \begin{tikzpicture}
    \matrix (m) [matrix of math nodes,row sep=2em,column sep=1em,minimum width=2em]
    {
      M & ~ & M\\
      ~& \R & ~\\};
    \path[-stealth]
    (m-1-1) edge node [above] {$\varphi$} (m-1-3)
    (m-1-1) edge node [left] {$\Psi_0$} (m-2-2)
    (m-1-3) edge node [right] {$\Psi$} (m-2-2);
  \end{tikzpicture}
\end{center}
The flux-functions are effectively related by precomposition with the inverse (pull-back)
\begin{align}
  \label{eq:frozen-in}
  \Psi = \Psi_0\circ \varphi^{-1} = {\varphi^{-1}}^*\Psi_0 =:\varphi_* \Psi_0
\end{align}
where we introduce the alias $\varphi_*:={\varphi^{-1}}^*$ for convenience. The magnetic one-forms are related in the same way, $d\Psi=d\varphi_*\Psi_0=\varphi_* d\Psi_0$, since the pull-back commutes with the exterior derivative, which corresponds to the usual frozen-in condition of ideal motion in the sense that the magnetic flux between pairs of advected points is preserved
\begin{align*}
  \Psi_0(b)-\Psi_0(a)=\int\limits_{\gamma} d\Psi_0
  = \int\limits_{\varphi(\gamma)} d\Psi = \Psi(\varphi(b)) - \Psi(\varphi(a)).
\end{align*}
This integral is commonly referred to as the \emph{poloidal magnetic flux} in toroidal confinement devices~\cite{dhaeseleer} or \emph{helical magnetic flux} in the vicinity of rational surfaces~\cite{zakharov-2008}.

If the diffeomorphism $\varphi$ is volume-preserving (coincidentally a symplectomorphism), $\varphi^*\omega=\omega$, then the Poisson-bracket is preserved. 
\begin{proof}
By direct computation, we have $(\varphi^*\{F,G\})\omega = (\varphi^*\{F,G\})(\varphi^*\omega)=\varphi^*(\{F,G\}\omega) =\varphi^*(dF\wedge dG) = (\varphi^*dF)\wedge (\varphi^*dG) = d(\varphi^*F)\wedge d(\varphi^*G) = \{\varphi^*F,\varphi^*G\}\omega$.
\end{proof}

Furthermore, if $\varphi$ is volume-preserving, the magnetic fields are related by push-forward. Indeed,
\begin{align*}
  i_B\omega = d\Psi = \varphi_*d\Psi_0
  =\varphi_*i_{B_0}\omega
  = i_{\varphi_*B_0} \varphi_*\omega,
\end{align*}
so $\varphi^*\omega=\omega \Rightarrow B = \varphi_* B_0$. This property is desirable in order to make contact with Faraday's law of induction, equation (\ref{eq:faraday}). In this context, volume-preserving diffeomorphisms (equally symplectomorphisms) represent smooth incompressible ideal motion.

The set of volume-preserving diffeomorphisms, denoted $S\Diff(M)$, has the structure of an infinite dimensional Lie group (Fréchet manifold). Arnold famously exploited this construct to interpret the Euler equations as geodesic equations~\cite{arnold-1966}. Ideal MHD has similar interpretations on semi-direct products~\cite{marsden-ratiu-weinstein,holm-marsden-ratiu,ono-1995,hattori-1994}.
We restrict our attention to the identity component $S\Diff^0(M)$, ruling out parity transformations for instance. The diffeomorphism $\varphi \in S\Diff^0(M)$, seen as a point on a manifold, is connected to the identity by a smooth path, namely a family of diffeomorphisms $\varphi_t\in S \Diff^0(M)$ for $t\in [0,1]$ with $\varphi_0=id$ and $\varphi_1=\varphi$. The family $\varphi_t$ generates, in turn, a smooth family of flux-functions via the frozen-in condition (\ref{eq:frozen-in})
\begin{equation}
  \label{eq:flux-functions}
  \Psi_t := \Psi_0 \circ \varphi_t^{-1} = {\varphi_t}_*\Psi_0.
\end{equation}

The \emph{variation} of $\varphi_t$ with respect to the parameter $t$ define a smooth family of vector fields on $M$ via
\begin{equation}
  \label{eq:eulerian-velocity}
  X_t := \partial_t \varphi_t \circ \varphi_t^{-1} : M\to TM
\end{equation}
called the \emph{Eulerian velocity field}. The picture is that the trajectory of a fluid element, initially at $p_0 \in M$, is $p(t) = \varphi_t(p_0)$, and its velocity is equal to the vector $dp/dt = \partial_t\varphi_t(p_0)= X_t(p(t))$. In local coordinates $p(t)=(x(t),y(t))$, this corresponds to the following system of ordinary differential equations
\begin{align*}
  \dot{x}(t) &= X^x(x(t),y(t),t), & \dot{y}(t) &=X^y(x(t),y(t),t).
\end{align*}
with $x(0)=x_0$ and $y(0)=y_0$ as initial conditions.

Because each diffeomorphism $\varphi_t$ is volume-preserving, the one-forms $i_{X_t} \omega$ are closed on $M$, indicating that every vector field $X_t$ is divergence-free.
\begin{proof}
$0=\partial_t \varphi_t^*\omega = \varphi_t^*\lie_{X_t} \omega$. Thus, $\lie_{X_t}\omega = di_{X_t}\omega = 0$, $\forall t$.  
\end{proof}

Consequently, by the Hodge decomposition theorem~\cite{schwarz}, $i_{X_t}\omega = dS_t + K_t dy$, where $S_t(x,y)$ are smooth single-valued potential functions, and $K_t$ are coefficients (constant on $M$).
In local coordinates, the Eulerian velocity field is expressed as
\begin{equation}
  \label{eq:xvol}
X_t = (\partial_y S_t + K_t) \partial_x -(\partial_x S_t)\partial_y
\end{equation}
which is automatically divergence-free. This corresponds to $\bm{X}_t=\curl[(S_t + K_t y) \nabla z]$ in the 3D slab picture. In local coordinates, the system of ODEs becomes
\begin{align}
  \dot{x}(t) &= \partial_yS(x(t),y(t),t)+K(t)\\
  \dot{y}(t) &= -\partial_xS(x(t),y(t),t).
\end{align}
The flux-function satisfies the following advection equation
\begin{align}
  \label{eq:advec-psi}
  \partial_t \Psi_t 
  = \{S_t + K_t y,\Psi_t\},
\end{align}
where $y$ is a locally-defined function such that $e^{iy}=w$ for $w\in S^1$. Note that the choice of $y$ does not affect the Poisson-bracket because different choices differ by a constant.
\begin{proof} By differentiating the frozen-in condition
  (\ref{eq:flux-functions}) with respect to $t$, we first show that
  \begin{align}
    \partial_t \Psi_t = -\lie_{X_t}\Psi_t.
  \end{align}
  Indeed, letting $p(t)=\varphi_t(p_0)$, the frozen-in condition reads $\Psi_0(p_0)=\Psi_t(p(t))$.  Differentiating with respect to $t$, applying the Leibniz and chain rules, we have
  \begin{align*}
    0 &= \partial_t\Psi_t(p) + d{\Psi_t}_p\left(\frac{dp}{dt}\right)
    = \partial_t\Psi_t(p) + d{\Psi_t}_p(X_t(p))\\
    &= \partial_t\Psi_t(p) + [\lie_{X_t}\Psi_t](p)
  \end{align*}
  Then,
\begin{align*}
  -\lie_{X_t}\Psi_t 
  &= -i_{X_t}d\Psi_t = -i_{X_t}i_{B_t}\omega = i_{B_t}i_{X_t}\omega \\
  &= i_{B_t}(dS_t + K_t dy)= \langle dS_t+K_t dy,B_t^\flat\rangle\\
  &= \star[(dS_t+K_tdy)\wedge\star(-\star d\Psi_t)] 
  \\&= \{S_t+K_t y,\Psi_t\}
\end{align*}
\end{proof}
Again because the diffeomorphisms $\varphi_t$ are volume-preserving, the family of poloidal magnetic fields is obtained by push-forward, $B_t = \varphi_{t*} B_0$. By differentiating this relation with respect to time, the following advection equation, recognised as Faraday's law of induction, is obtained
\begin{align}
  \label{eq:faraday}
\partial_t B_t =- [X_t,B_t].
\end{align}
In local Cartesian coordinates, this reads as $\partial_t \bm{B} = - \bm{X}\cdot\nabla\bm{B}+\bm{B}\cdot\nabla\bm{X}= \curl(\bm{X}\times \bm{B})$ where the last step follows from the fact that both fields are divergence-free.

\begin{remark}
  Unless $\varphi_t$ is an isometry (translation and rotation), the Laplacian $\Delta \Psi_t$, is not an advected quantity.
  %
\end{remark}

\subsection{Flow-maps and advection through $t$-independent potential
  functions}
In the case where the potential functions $S_t=S$ and coefficients $K_t=K$ are independent of $t$, the family of diffeomorphisms $\varphi_t$ coincides with the flow-map of the corresponding fixed vector field $X$ on $M$. The solution to the advection equation (\ref{eq:advec-psi}) can be formally computed as
\begin{align}
  \Psi_t &= \exp(t\{H,.\})\Psi_0\nonumber\\
  &= \Psi_0 + t\{H,\Psi_0\} + \frac{t^2}{2}\{H,\{H,\Psi_0\}\} + \ldots\label{eq:liouv-exp}
\end{align}
where $H = S+Ky$. The potential function $H$ generates a near-identity (canonical) transformation. Evaluated at any point $p=(x,y)$, such expansion is analytic in $t$ when it converges.

\begin{figure}
  \centering
  \includegraphics[width=\columnwidth]{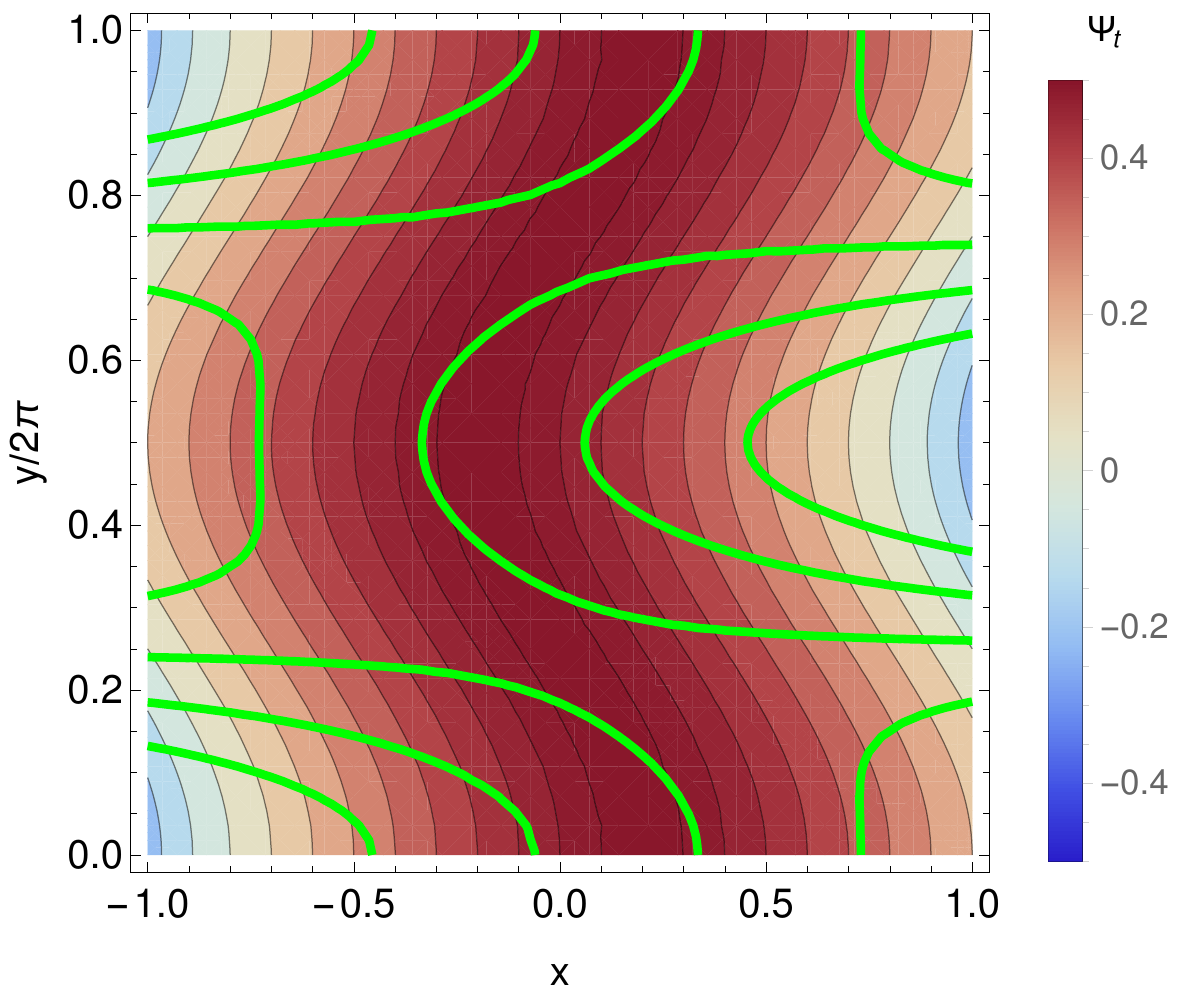}
  \caption{Advected flux $\Psi_t(p) =\Psi_0[\varphi_t^{-1}(p)]$ generated by the $t$-independent potential function $S(x,y) = \sin y$ for $t=0.2$ from initial HKT configuration $\Psi_0(x,y)=\tfrac{1}{2}(1-x^2)$. The colour contours represent the level sets of the flux-function and the thick green curves are level sets of the Laplacian $\Delta \Psi_t$. The fact that the two do not overlap indicates lack of force balance.}
  \label{fig:advec-siny}
\end{figure}
For example, consider the initial HKT configuration $\Psi_0(x,y)=\tfrac{1}{2}(1-x^2)$. We study the smooth ideal motion generated by the $t$-independent potential function $S(x,y)=\sin y$ and $K=0$. The corresponding system of ODEs for the coordinates has solution
\begin{align*}
  &  \left|
  \begin{array}{l}
    \dot{x} = \cos y\\
    \dot{y} =0
  \end{array}
\right.&\iff&&
& \left|
\begin{array}{l}
  x(t) = x_0+t\cos y_0\\
  y(t) = y_0
\end{array}
\right.
\end{align*}
The advected flux-function is thus
\begin{align*}
  \Psi_t(x,y)& = \Psi_0(x-t\cos y,y) = \frac{1}{2}\left[1-(x-t\cos y)^2\right]\\
  &= \Psi_0(p) +t x\cos y - \frac{1}{2}t^2 \cos^2y
\end{align*}
which matches the expansion in equation (\ref{eq:liouv-exp}) with $\{\sin y,\Psi_0\} = -\tfrac{1}{2}\{\sin y , x^2\} = x\cos y$, $\{\sin y,x\cos y\} = -\cos^2y$ and $\{\sin y,\cos^2 y\}=0$. The level sets of the advected flux-function are shown on Figure \ref{fig:advec-siny} for $t=0.2$. The flux-levels remain unbroken for all values $t$ by virtue of the frozen-in condition. This configuration is not in force balance since these level sets of the flux-function do not align with that of its Laplacian,
\begin{equation*}
  \Delta \Psi_t = 1 + t x \cos y - t^2\cos(2y),
\end{equation*}
depicted by the green thick lines on Figure \ref{fig:advec-siny}. The obstruction occurs at linear order in $t$, with the left-hand side of equation (\ref{eq:force-balance-diff}) being
\begin{align*}
  \{\Delta\Psi_t, \Psi_t\}   =-t x^2\sin y + O(t^2).
\end{align*}

\begin{figure}
  \centering
  \includegraphics[width=\columnwidth]{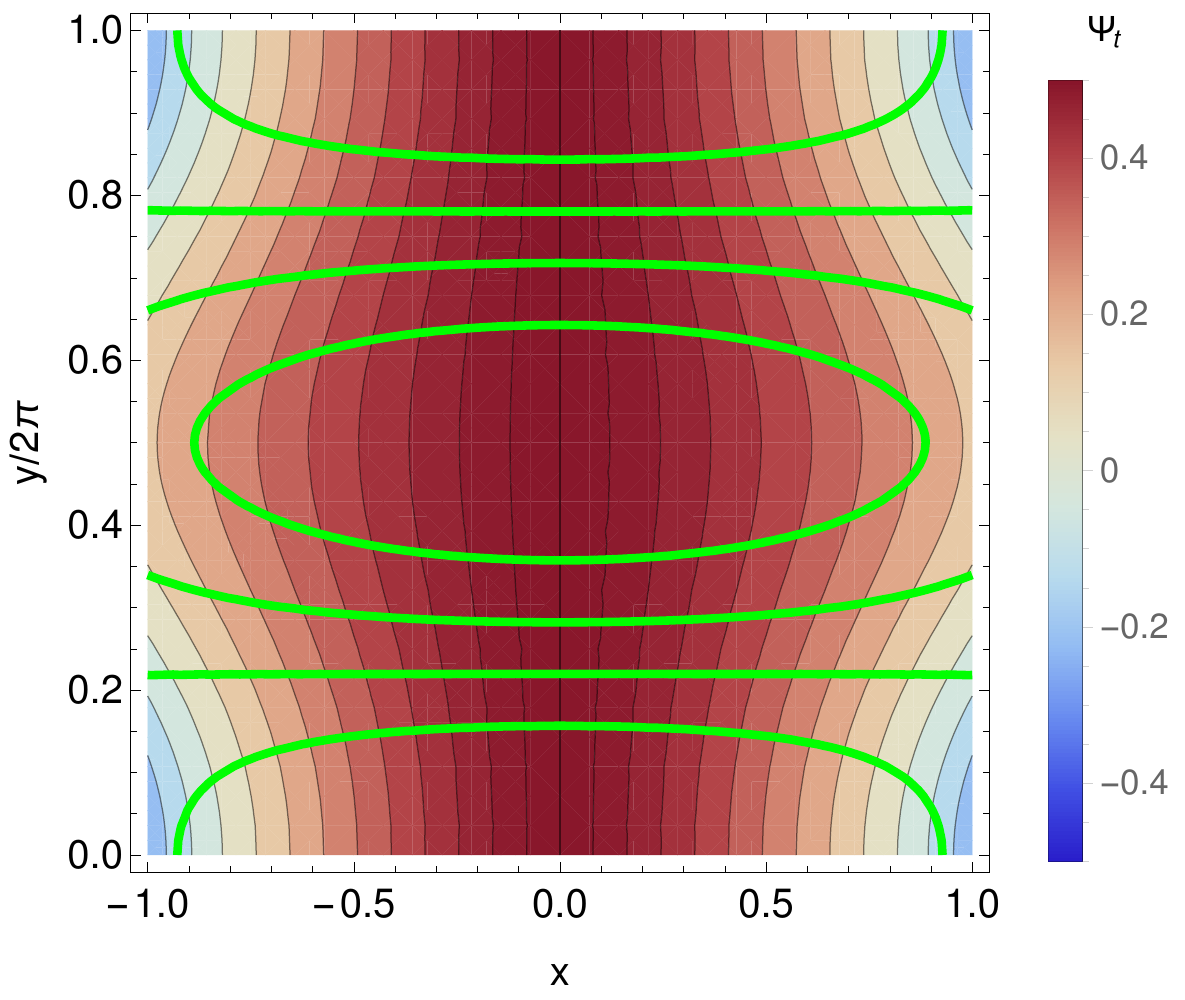}%
  \caption{Same caption as Figure \ref{fig:advec-siny} but for
    generating potential function $S(x,y) =-x \sin y$.}
  \label{fig:advec-xsiny}
\end{figure}
A more intricate example is the flow-map generated by the potential function $S(x,y)=-x\sin y$ and $K=0$. The reader may verify that the coordinate transformation satisfies
\begin{align*}
  &  \left|
  \begin{array}{l}
  \dot{x} = -x\cos y \\
  \dot{y} =\sin y
  \end{array}
\right.&\iff&&
&\left|
  \begin{array}{l}
  x(t) = x_0(\cosh t-\sinh t \cos y_0)\\
  y(t) = 2\tan^{-1}\left[e^t\tan\left(\tfrac{y_0}{2}\right)\right]
\end{array}
\right.
\end{align*}
Rather neatly, the inverse map is obtained by negating the $t$-parameter such that the advected flux becomes
\begin{align*}
  \Psi_t(x,y)%
  &= \frac{1}{2}\left[1-x^2\left(\cosh t + \sinh t \cos y\right)^2\right].
\end{align*}
It can be verified that the Taylor series in $t$ around $t=0$ of this flux-function matches the result from the expansion in equation (\ref{eq:liouv-exp}). The level sets are shown on Figure \ref{fig:advec-xsiny} for $t=0.2$. The flux-levels remain unbroken for all values $t$ by the advection equation. This particular choice of potential function $S(x,y)$ preserves the even parity of the flux-function $\Psi_t(-x,y)=\Psi_t(x,y)$. This configuration is however not in force balance since the level sets of the flux-function do not align with that of its Laplacian, depicted by the green thick lines on Figure \ref{fig:advec-xsiny}. The obstruction occurs again at linear order in $t$ with the left-hand side of equation (\ref{eq:force-balance-diff}) being equal to
\begin{align*}
  \{\Delta\Psi_t, \Psi_t\}   =-tx(2-x^2)\sin y + O(t^2).
\end{align*}

These examples suggest that the reduced MHD equilibrium problem be viewed as the process of tuning the potential functions $S_t(x,y)$ until the level sets of the generated flux-functions and their Laplacian overlap for all $t$. In many respects, this is the point of view adopted by the Variational Moments Equilibrium Code (VMEC)~\cite{vmec}, where three-dimensional magnetic configurations are obtained by minimising $\bm{J}\times\bm{B}-\nabla p$ through a steepest-descent method on the Fourier coefficients parametrising the (pre-existing) nested flux-surfaces. However, in the case of mirror-symmetric boundary conditions, it is proven in the next section that cancelling the force terms order by order in $t$ leads to the vanishing of the entire family of potential functions $S_t$. The initial configuration is thus said to be \emph{rigid} against ideal MHD motion respecting the mirror symmetry of the flux-function.

\section{Hahm-Kulsrud-Taylor (HKT) problem}
\label{sec:hkt}

Let us consider the initial flux-function $\Psi_0(x,y) = \frac{1}{2}(1-x^2)$ with straight flux-levels. The configuration
\begin{align*}
  d\Psi_0 &= -x dx, &
  \star d\Psi_0 &= -xdy, &
B_0&=x\partial_y, &
\Delta \Psi_0 &= 1,
\end{align*}
satisfies the force balance condition and represents an initial MHD equilibrium. We wish to find a family of volume-preserving diffeomorphisms $\varphi_t$ from the identity $\varphi_0=id$ to $\varphi_1=\varphi$ such that the force balance condition is achieved at every $t$,
\begin{align}
\label{eq:force-balance}
\{\Delta\Psi_t,\Psi_t\}=0,\quad \forall t \in [0,1].
\end{align}
%


Differentiating (\ref{eq:force-balance}) with respect to $t$ and evaluating at $t=0$ yields
\begin{align}
  \{\Delta\Phi,\Psi_0\} + \cancel{\{\Delta\Psi_0,\Phi\}} = 0 \Rightarrow
\{\Delta\Phi ,x^2\}=0 \Rightarrow \Delta \partial_y\Phi = 0 \label{eq:harm}
\end{align}
where $\Phi := \partial_t\Psi_t\big|_{t=0} = -d\Psi_0(X_0)$ is the initial \emph{rate of change} of $\Psi_t$ along the path $\varphi_t$. Force-balance requires this quantity to be of the form
\begin{equation}
  \label{eq:phi_sol}
  \Phi = h(x,y) + f(x)
\end{equation}
where, by standard separation of variables,
\begin{multline}
  \label{eq:harmonic}
  h(x,y)=\sum_{n=1}^\infty [A_n\cosh(nx)\cos(ny) + B_n\cosh(nx)\sin(ny) \\
  + C_n\sinh(nx)\cos(ny) + D_n\sinh(nx)\sin(ny)]
\end{multline}
is a harmonic function on $M$ with $A_n$, $B_n$, $C_n$ and $D_n$ determined by boundary conditions.

The advection equation (\ref{eq:advec-psi}) states that the rate of change is related to the potential function at $t=0$ via
\begin{align}
\label{eq:phirelates}
  \Phi =& \{S_0+K_0 y,\Psi_0\} = x (\partial_y S_0 + K_0).
\end{align}
Equating this with (\ref{eq:phi_sol}), the potential function is found to satisfy
\begin{align}
  \label{eq:sfromphi}
  \partial_y S_0 
  = \frac{h}{x} + \frac{f}{x} -K_0.
\end{align}
Since $\oint \partial_y S_0 dy = 0$ and $\oint h dy = 0$, the function $f$ is seen, after integrating equation (\ref{eq:sfromphi}) over $y$, to be of the form $f(x)=K_0 x$. The initial rate of change is thus a purely harmonic function $\Phi=h+K_0 x$ and the potential function has the form
\begin{align}
  S_0(x,y) =& \frac{H(x,y)}{x} + g(x)
\end{align}
where
\begin{multline}
  H(x,y)=\sum_{n=1}^\infty \frac{1}{n}[A_n\cosh(nx)\sin(ny)-B_n\cosh(nx)\cos(ny)\\
  +C_n\sinh(nx)\sin(ny) - D_n\sinh(nx)\cos(ny)].
\end{multline}
We note that the $g(x)$ term does not contribute to displacing flux-levels at $t=0$, since $dg(B_0)=x\partial_y g=0$. The potential function above leads to the following Eulerian velocity field
\begin{align}
  \label{eq:X0}
  X_0 =&\left(\frac{h}{x}+K_0\right)\partial_x +\left(\frac{H-x\partial_x H}{x^2} + g' \right)\partial_y.
\end{align}

Equation (\ref{eq:phirelates}) states that the initial rate of change vanishes on the neutral line, $\Phi(0,y) = 0$. This requirement implies $A_n=0$ and $B_n=0$, which is a necessary condition for smooth incompressible ideal motion. If $A_n$ and $B_n$ were non-zero, the potential function and the Eulerian velocity field would become singular. The boundary conditions on $\Psi_t$ must thus be compatible with the fact that the initial rate of change $\Phi$ can only be made an odd function of $x$.

In fact, we prove by strong induction that it is impossible to deform the plasma such that the mirror symmetry of the flux-function is preserved, $\Psi_t(-x,y)=\Psi_t(x,y)$, for all $t$. The base case consists of the fact that if mirror symmetry were to be preserved, $C_n=0$, $D_n=0$ and $K_0=0$, namely the initial rate of change vanishes $\Phi= \partial_t\Psi_t\big|_{t=0}=:\Psi_0^1 = 0$. Let us assume that mirror symmetry of the flux-function is respected up to its $i^{th}$-derivative in $t$ at $t=0$, i.e.  $\Psi_0^j:=\partial^{(j)}_t\Psi_t\big|_{t=0}= 0$ for $j=1,\ldots i $. The force-balance condition gives, by virtue of the general Leibniz rule,
\begin{align*}
  \sum_{j=0}^{i+1} \binom{i+1}{j}  
  \{ \Delta \Psi_0^{i+1-j},\Psi_0^j\}& = 0 \Rightarrow \{\Delta \Psi_0^{i+1},\Psi_0\} =0 \Rightarrow\\
  \Delta \partial_y \Psi_0^{i+1} &= 0,
\end{align*}
i.e. $\Psi_0^{i+1} = h^{i+1}(x,y) + f^{i+1}(x)$, similarly to equation (\ref{eq:phi_sol}). Differentiating the advection equation (\ref{eq:advec-psi}) $i$ times and evaluating at $t=0$, one also obtains that
\begin{align}
  \label{eq:ithpsirelates}
  \Psi_0^{i+1}& =\sum_{j=0}^i\binom{i}{j} \{S_0^{i-j}+K_0^{i-j}y,\Psi_0^j\} \nonumber \\
  &= x(\partial_y S_0^i + K_0^i)
\end{align}
where $S_0^j := \partial^{(j)}_t S_t\big|_{t=0}$ and $K_0^j:=\partial^{(j)}_tK_t\big|_{t=0}$. Integrating over $y$, the function $f^{i+1}$ must be of the form $f^{i+1}(x)=K_0^i x$, the only even-parity preserving choice however being $K_0^i=0$.
As before, equation (\ref{eq:ithpsirelates}) provides the boundary condition that $\Psi_0^{i+1}(0,y) = 0$, thereby eliminating all even-parity terms from the harmonic function $h^{i+1}$. We thus conclude that $\Psi_0^{i+1} = 0$. By induction, this result extends to all derivatives of the flux-function with respect to $t$ at $t=0$.

If there exists a smooth family of diffeormophisms such that the advected flux-function remains even in $x$ and in force balance for all values of the time-like parameter $t$, then it is non-analytic in $t$ at $t=0$. Conversely, a path of diffeormorphisms $\varphi_t$, analytic in $t$ at $t=0$ and such that $\Psi_t$ is in force balance, does not preserve the even-parity of the flux-function. In particular, mirror-symmetric boundary conditions such as
\begin{equation}
  \label{eq:bceven}
  \Psi_t(\pm a,y) = \Psi_0(a)[1 + t^i \cos (n y+\alpha)]
\end{equation}
are \textbf{unsupported} for all $t$ and all integer power $i \geq 1$ ($\forall n\in \mathbb{N}$, $\forall \alpha \in [0,2\pi]$). Consequently, the time-like variable $t$ cannot serve as an expansion parameter to control the application of mirror-symmetric boundary perturbations. In other words, the HKT initial equilibrium is rigid to smooth ideal deformations that preserve the even property of the flux-function with respect to the neutral line. 

The above result does not rule out the possibility of connecting the initial HKT configuration to a parity-preserving state with curved level sets via a path of diffeomorphisms that is non-analytic with respect to the control parameter $t$. Such ideal motion would be quite special from the point of view of a continuous deformation of the boundary. In fact, formulated as an over-determined boundary value problem $-\Delta \Psi=V'(\Psi)$ with $d\Psi=0$ at $x=0$ and $\Psi=0$ on the edge of an mirror-symmetric domain in $x$, the existence (and regularity) of solutions with non-straight level sets is rather unlikely~\cite{serrin-1971}. The details of this question will be addressed in future work.

\subsection{Flow-map of equilibrium-preserving generating potential function}
\begin{figure}
  \centering
  \includegraphics[width=\columnwidth]{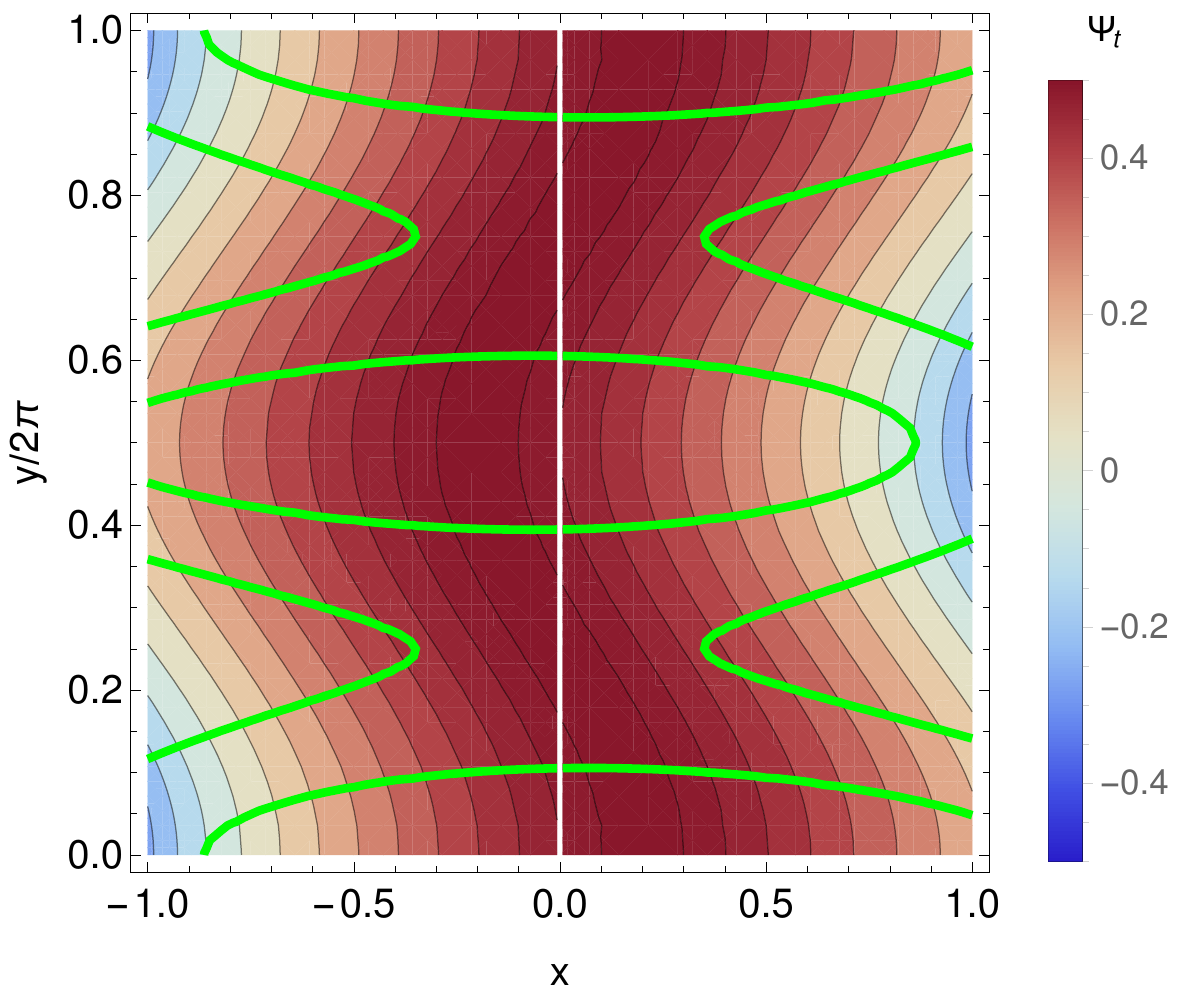}%
  \caption{Advected flux $\Psi_t(p)$ generated by the $t$-independent potential function $S(x,y) = \sinh x \sin y/x$ for $t=0.2$ from initial HKT configuration $\Psi_0(x,y)=\tfrac{1}{2}(1-x^2)$, computed using equation (\ref{eq:liouv-exp}) up to fifth order. The colour contours represent the level sets of the flux-function and the thick green curves are level sets of the Laplacian $\Delta \Psi_t$. The fact that the two do not overlap indicates lack of force balance. The obstruction is however quadratic in $t$, instead of linear as in figures \ref{fig:advec-siny} and \ref{fig:advec-xsiny}.}
  \label{fig:advec-sinhxOxsiny}
\end{figure}
A typical example of a smooth generating potential function respecting force-balance at $t=0$ would be
\begin{align*}
  S_0(x,y) = \frac{\sinh x}{x} \sin y.
\end{align*}
Using this potential function in equation (\ref{eq:liouv-exp}) to compute the flow-map of a $t$-independent vector field, the advected flux-function is reconstructed as
\begin{multline*}
  \Psi_t = \Psi_0 + t \sinh x \cos y \\
  + t^2\frac{\sinh x(\sinh x \sin^2y-x\cosh x)}{2x^2} + O(t^3),
\end{multline*}
whose level sets are shown in Figure \ref{fig:advec-sinhxOxsiny}. While the green lines, depicting level sets of the Laplacian $\Delta \Psi_t$, still do not align with the flux-levels, the obstruction to force-balance occurs now at $O(t^2)$, as can be checked by computing $\{\Delta \Psi_t,\Psi_t\}$. This suggests that an equilibrium-preserving diffeomorphism will have to be generated by time-dependent potential functions. In other words, smooth ideal incompressible motion from an initial equilibrium to another accessible state cannot be in any one-parameter group of diffeomorphisms. This is to be expected by the non-linearity of the "algebraic" force-balance condition in contrast to the advection equation.

\section{Discussion}
\label{sec:conclusion}
In this paper, we addressed the problem of accessing MHD equilibrium states via smooth ideal motion and highlighted the main obstruction to finding a smooth family of force-balanced flux-functions in the simplest two-dimensional slab case known as the Hahm-Kulsrud-Taylor problem. The MHD force-balance condition was conveniently written as the Poisson-commutation of the flux-function $\Psi_t$ with its Laplacian $\Delta \Psi_t$. This form can be achieved whenever a Grad-Shafranov equation holds (symmetry by isometry). The strategy was then to find a family of volume-preserving diffeomorphisms to advect an initial equilibrium configuration into another with deformed but topologically-equivalent flux-levels. The use of volume-preserving maps (symplectomorphisms) insured invariance of the Poisson-bracket, as well as advection of the magnetic field (Faraday's law). The problem was then to solve for the family of generating potential functions $S_t$, which are in effect $t$-dependent Hamiltonian functions for the Eulerian velocity field $X_t$. Starting with the one-dimensional HKT equilibrium configuration, the initial rate of change $\Phi = \partial_t\Psi_t|_{t=0}$ of the flux-function was found to be a harmonic function. However, through its relation to the generating function, the initial rate of change had to vanish on the line of critical points of $\Psi_0=\frac{1}{2}(1-x^2)$ at $x=0$. This condition excluded the application of boundary conditions that retain the mirror symmetry of the flux-function. It was proved by strong induction on the Taylor expansion in $t$ that those boundary conditions had to be excluded at all orders. 

This result showed that, within the class of smooth solutions and via ideal motion (perfect advection of flux), MHD equilibria do not deform when subject to certain boundary perturbations. This rigidity implies from a computational and numerical point of view that at least one of the constraints on the motion must be relaxed in order to generate solutions, e.g. allowing for discontinuities, including the effect of finite resistivity (dissipate flux accumulation), enabling regions of force imbalance, etc. Deciding which workaround is more suitable is left for debate.

It is well-understood that there always is a finite amount of resistivity in a physical system, that allows for profile smoothing and release of flux in infinitesimal regions around resonant surfaces (neutral line in the flux-function). This is reminiscent of d'Alembert's paradox in fluid dynamics, where the drag force vanishes in the limit of an inviscid potential flow, contradicting the phenomenological observation of substantial drag at high Reynolds numbers.

Without resistivity, the commonly accepted solution to the HKT problem is a boundary layer treatment originally proposed by \citet{rosenbluth-1973} for the non-linear saturation of the internal kink instability. In a small region around the neutral line, a discontinuous solution is obtained with approximate force balance, which is then asymptotically matched with the ideal solution on the outside. This method is, in light of this paper, not to be mistaken with a strictly ideal treatment.

\bibliography{biblio}

\appendix

\section{Proof of Jacobi identity of the Poisson Bracket}
\label{sec:proof-jacobi}
By virtue of the Cartan formula, the identity $i_{[X,Y]} = [\lie_X,i_Y]$ and the divergence-free property of Hamiltonian vector fields, we have
  \begin{align*}
    i_{[X_F,X_G]}\omega &= [\lie_{X_F},i_{X_G}]\omega
    = \lie_{X_F}i_{X_G}\omega - i_{X_G}\cancel{\lie_{X_F}\omega}\\
    &= i_{X_F}\cancel{d i_{X_G}\omega} + di_{X_F}i_{X_G}\omega\\
    &= d(\{F,G\}) = i_{X_{\{F,G\}}}\omega.
\end{align*}
Hence, the following identity holds
  \begin{align}
    \label{eq:vec-poisson}
    X_{\{F,G\}} = [X_F,X_G].
  \end{align}
The Jacobi identity for the Poisson bracket then follows from the Jacobi identity satisfied by the Lie bracket of vector fields
\begin{align*}
X_{\{F,\{G,H\}\} + \text{cycl}} 
  = [X_F,[X_G,X_H]] + \text{cycl} = 0.
\end{align*}

\end{document}